%Paper: hep-lat/9209010
%From: LOMBARDO@CPWSCA.PSC.EDU (Maria-Paola Lombardo)
%Date: Fri, 4 Sep 1992 14:50:26 -0400 (EDT)

\magnification=1000
\baselineskip=16pt
\pageno=0
\rightline{ILL-(TH)-92-\# 18}
\rightline{SEPTEMBER 1992}
\rightline{CERN-TH.6630/92}
\vskip1truecm

\centerline{ UNIVERSAL PROPERTIES OF CHIRAL SYMMETRY BREAKING }
\vskip2truecm

\centerline{Aleksandar KOCI\' C\footnote{$\,^{ }$}{ILL-(TH)-92-18} }
\centerline{\it Theory Division, CERN, CH-1211 Geneva 23,
Switzerland}

\vskip1truecm

\centerline{John B. KOGUT and Maria-Paola LOMBARDO$^{*}$
\footnote{$\,^{ }$}{CERN-TH.6630/92}}
\centerline{\it Department of Physics, University of
Illinois at Urbana-Champaign}
\centerline{ \it 1110 West Green Street, Urbana,
IL 61801-3080 USA}

\vskip2truecm

\centerline{\bf Abstract}

{\narrower
We discuss chiral symmetry breaking critical points from the perspective
of PCAC, correlation length
scaling and the chiral equation of state. A scaling theory for the
ratio $R_\pi$ of the pion to sigma masses is presented.
The Goldstone character of the pion and properties of the longitudinal and
transverse chiral susceptibilities determine
the ratio $R_\pi$ which can be used to locate
critical points and measure critical indices such as
$\delta$.We show how PCAC and correlation length
scaling determine the pion mass' dependence on the chiral condensate
and lead to a practical method to measure the anomalous dimension
$\eta$. These tools are proving useful in studies of the chiral
transition in lattice QED and the quark-gluon plasma transition in
lattice QCD.}

\footnote{$\,^{ }$}{September 1992}
\footnote{$\,^{ }$}{$^*$ On leave from Infn, Sezione di Pisa, Italy}
\smallskip
\vfill\eject

\noindent
1. INTRODUCTION
\vskip5truemm

Chiral symmetry phase transitions are particularly interesting because
of their physical relevance and their theoretical constraints. The order
parameter for the high temperature quark-gluon transition of QCD is the
chiral condensate and considerable effort both in phenomenological studies
and computer simulations are underway [1].
Non-compact lattice QED experiences
a chiral transition at strong coupling and one needs new practical theoretical
tools to understand the character of this transition [2]. Four Fermi models
with N species of fermions in less than four dimensions have chiral
transitions, and 1/N expansions indicate that these transitions
correspond to ultra-violet fixed points of the renormalization group
[3], [4].
%[WILSON,ROSENSTEIN ET AL., HKK].
These models illustrate the intimate relation between renormalizability,
compositeness and hyperscaling in four Fermi and Yukawa models. Composite
models of the Higgs mechanism which underlies the Standard Model typically rely
on a chiral symmetry breaking phase transition to make composite mesons
which can produce heavy W and Z particles.
Although the original versions
of this idea (Technicolor) copied the asymptotically free dynamics of
QCD, other versions imagine theories which are strongly coupled at short
distances [5].
%[T. APPELQUIST, Dynamical EW Symmetry Breaking, YCTP-P23-91].
Interesting models of a heavy top quark based on gauged Nambu
Jona Lasinio models have been considered [6].
%[MIRANSKY ET AL., BARDEENN HILL LINDNER].
In this paper we shall be
particularly interested in models where chiral symmetry breaking is
a short distance phenomenon and we shall concentrate on non-asymptotically
free dynamics.

Chiral symmetry breaking phase transitions have important simplifying
features which will allow us to study them in several
complementary ways. Our theoretical framework will be that of statistical
mechanics and concepts borrowed from ferromagnetic phase transitions
such as the equation of state, scaling variables, correlation length
scaling and hyperscaling relations between critical indices, etc. will
be used heavily here. This 'standard approach' gains far greater
predictive power when current algebra relations following from the
underlying continuous chiral symmetry of the system's Lagrangian
are developed. The most familiar deduction of this approach is
the Goldstone character and low energy theorems concerning the pion.
But we shall present additional equally firm results here involving
the chiral partner, the sigma. One of the goals of this work is to
develop practical methods to extract critical indices from chiral
symmetry breaking transitions and classify critical points as Gaussian
(trivial) or non-Gaussian (nontrivial). We shall see that the ratio
$R_\pi=M_\pi^2/M_\sigma^2$ and its dependence on coupling and bare fermion
mass predicts the critical coupling and the the critical index
$\delta$. We shall relate $R_\pi$ to the transverse and longitudinal
susceptibilities of the chiral phase transition and show that $R_\pi$
is completely determined by the theory's equation of state. The
resulting formula has already proved its usefulness in studies of lattice
QED [N=2] and we anticipate applications to the other models listed
above. Most past work on lattice QED and its chiral symmetry breaking
phase transition have concentrated on the order parameter
$<\bar\psi\psi>$ and its equation of state [7]. Since these quantities
are measured at nonvanishing bare fermion mass, the critical point must
always be inferred and a rapid crossover between a symmetric and a
broken phase occurs in raw simulation data. This obscures the
critical point and the value of critical indices. However, since
$R_\pi$ is determined by the same equation of state, any hypothesis
one makes concerning $<\bar\psi\psi>$ and its equation of state must
predict $R_\pi$ or be discarded. This fact should help us find the correct
theoretical ideas describing the phase transition of lattice QED, and
should be applicable to other models as well.

Following the same philosophy, we shall study $M_\pi$ as a function
of $<\bar\psi\psi>$ itself. We shall see that combining correlation
length scaling with the Goldstone description of the pion, yields
particularly effective methods to find the critical coupling and the
anomalous dimension $\eta$. Since $M_\pi$ and $<\bar\psi\psi>$
are the easiest observables to measure in simulations and typically
the least technical quantities to calculate in models, this approach
to $\eta$ should prove quite useful. In addition, the anomalous dimension
$\eta$ is central to discussions of nontriviality and thus lies at the
heart of the issue of whether non-asymptotically free theories exist.

The rest of this paper is organized as follows. In Sec.2 we introduce
our general theoretical framework and discuss renormalization group
trajectories, the chiral equation of state and the physics behind
hyperscaling. In Sec.3 we concentrate on $R_\pi$ and show that it is
determined by the chiral equation of state and can be used, because
of the Goldstone nature of the pion, to locate the critical coupling
and measure the critical index $\delta$ in an (almost) model
independent fashion. Sec.4 considers the chiral equation of state
and the constraints that PCAC places on it. In Sec.5 the relation
of the pion mass, and the chiral condensate is developed resulting
in a novel way to find the critical point and the anomalous
dimension $\eta$. Sec.6 considers the fermion mass and the index
$\delta$ and discusses the scaling region where the analysis
of this paper applies. It also discusses non-asymptotically free
dynamics and the breaking of scale invariance at the
critical point. In Sec.7 we
illustrate some of our analyses in the context of two models, the
four-fermi and the linear sigma model. Sec.8 contains some brief
concluding remarks.

\vskip1truecm

\noindent
2. RENORMALIZATION GROUP TRAJECTORIES AND HYPERSCALING
\vskip5truemm

The general idea of renormalizability is that a theory's  cutoff dependence
can be absorbed
into a finite set of bare parameters in such a way that its low energy physics
is insensitive to the cutoff. Once this is done, it is possible to
find the lines of constant physics (RG trajectories)
in the bare parameter space. These lines are uniquely defined no matter
what observable is taken.
In this way low-energy quantities depend on each
other and not on the cutoff.
We will see that this is possible if hyperscaling
is obeyed.

In our discussion we shall assume that the reader is
familiar with traditional statistical mechanics topics of homogeneity,
the Equation of State (EOS) of ferromagnets, correlation length
scaling etc. We shall pass between the languages of field theory
and statistical mechanics freely in much the same spirit as the
textbook by C. Itzykson and J.-M. Drouffe [8].
%[ID].

The hyperscaling hypothesis claims that the only relevant scale in
the critical region is the macroscopic correlation length
$\xi$. If this hypothesis holds, it is
possible to do dimensional analysis using this correlation length as a scale.
As a consequence, all dimensionless observables, e.g. mass ratios and
renormalized couplings, should be independent of $\xi$.
The hyperscaling hypothesis can be stated as

$$
F_{sing}=t^{2-\alpha} F(h/t^\Delta)\sim \xi^{-d}
\eqno(2.1)
$$
where notation is standard (F is the free energy, h is the external
magnetic field, t is the deviation of the dimensionless
temperature (coupling) from the critical point, etc.).
All thermodynamic quantities are obtained by taking
the derivatives of the free energy. In particular, the
order parameter, defined as $<\phi>=\partial F_{sing}/\partial h$,
satisfies the Equation of State (EOS)

$$
<\phi>=t^{2-\alpha -\Delta}F'(h/t^\Delta) =t^\beta F'(h/t^\Delta)
\eqno(2.2)
$$
This relation also defines the magnetic exponent: $\beta=2-\alpha-\Delta$.
Similarly, the susceptibility exponent, $\gamma$, is obtained from

$$
\chi=
{{\partial <\phi>}\over{\partial h}}
= t^{\beta-\Delta} F''(h/t^\Delta) \equiv
t^{-\gamma} F''(h/t^\Delta)
\eqno(2.3)
$$
i.e. $\gamma=\Delta-\beta$.

For a given channel, the corresponding correlation length (= inverse mass)
is defined by

$$2d\xi_\phi^2={{\int_x |x|^2<\phi(x)\phi(0)>}\over
{\int_x<\phi(x)\phi(0)>}}
\eqno(2.4)
$$
If the field $\phi$ develops a vacuum expectation value, then connected
correlation functions should be taken in eq.(2.4).
This will be understood throughout the paper.
In the scaling region its behavior is given by

$$
\xi=t^{-\nu} g(h/t^\Delta)
\eqno(2.5)
$$
Combined with eq.(2.1), it leads to the hyperscaling relation between the
critical exponents: $d\nu=2-\alpha$. The other relations between the
exponents follow from the scaling form of the two-point function:
$<\phi(x)\phi(0)>=1/|x|^{2d_\phi}f(x/\xi)$ [9].
%[Ka].

Since hyperscaling is an important concept, we should explain its meaning and
outline possible implications.
It is generally believed that violations of hyperscaling lead to
triviality. This is due to various inequalities between
certain combinations of critical
indices [10].
%[B,S].
In simple models like scalar field theories and spin
systems, the quantity that measures the violation of hyperscaling is
the dimensionless
renormalized coupling [11].
%[B,P].
It is defined through the nonlinear susceptibility $\chi^{(nl)}$

$$
g_R=-{ {\chi^{(nl)}}\over{\chi^2 \xi^{d}} }, \,\,\,\,\,\,
\chi^{(nl)}={ {\partial^3 <\phi>}\over{\partial h^3} }
\eqno(2.6)
$$
The renormalized coupling is essentially the properly normalized connected
four-point function. In terms of correlation functions, the
nonlinear susceptibility is the zero-momentum projection of the four-point
function

$$
\chi^{(nl)} = \int_{xyz}<\phi(x)\phi(y)\phi(z)\phi(0)>_{conn}
\eqno(2.7)
$$
The $\xi^d$ term in the denominator of eq.(2.6) is to account for an extra
integration. By differentiating
the free energy and using the EOS for the correlation length
(eq.(2.5)), we arrive at

$$
g_R = t^{-2\Delta+\gamma+d\nu} H(h/t^\Delta)
\eqno(2.8)
$$
We can trade $t$ for the correlation length in eq.(2.4) and rewrite
eq.(2.8)
in terms of $\xi$ as

$$
g_R = \xi^{(2\Delta-\gamma-d\nu)/\nu} {\tilde H}(h/t^\Delta)
\eqno(2.9)
$$

In specific (ferromagnetic) models an estimate of the exponent can be made:
since multi-spin correlations can not extend over a larger range than pair
correlations, the following inequality holds [10]
%[B,S]
$$
2\Delta\leq \gamma +d\nu
\eqno(2.10)
$$
Because $g_R$ is dimensionless, hyperscaling implies that it is
independent of $\xi$ and the exponent
in eq.(2.9) must vanish. In this case the renormalized
coupling is a function of only one bare variable.
Clearly, strict inequality in eq.(2.10)
implies triviality. We mention that an equivalent way of stating the
above inequality is $d\nu\geq 2-\alpha$ [12].
%[P].
It means that the singular part
of the free energy vanishes at most as fast as required by hyperscaling i.e.
$F_{sing}\geq \xi^{-d}$.

In a similar fashion the scaling of the mass ratios can be derived [13].
%[K]
If hyperscaling is satisfied, the ratio of
any particular pair of masses satisfies

$$
R(t,h)=G(h/t^\Delta)
\eqno(2.11)
$$
where $G(y)$ is a universal function. Comparing it with the expression for
the renormalized coupling, we see that both observables depend on the same
variable. One of the relations can be inverted to solve for the bare variable
e.g. $h/t^\Delta=H^{-1}(g_R)$.
This defines an RG trajectory for each value of $g_R$
and can be used to obtain the relation between the two observables
$R=R(g_R)$. Note that this relation is independent of the bare parameters.
The same manipulation can be done with two mass ratios. We note
that the important point in the inversion is that both observables depend
on just one bare variable, so that the inverse relation can always be found,
at least in some regions of parameter space.
\vskip1truecm

\noindent
{3. $\sigma-\pi$ SPLITTING, CRITICAL EXPONENTS AND EQUATION OF STATE}
\vskip0.5truecm

In this section we will relate the $\sigma-\pi$ splitting with other
universal quantities like critical indices and universal amplitude ratios.
Since the scalar and pseudoscalar propagators represent fluctuations of
the order parameter, information about the $\sigma-
\pi$ spectrum is contained in the equation of state.
Before we consider the derivation, we briefly illustrate the connection
between the behavior of $<\bar\psi\psi>$ and the $\sigma-\pi$ splitting.
We use the spectral representation to express the order parameter

$$
<\bar\psi\psi> =\int_{-\infty}^{+\infty} d\lambda \rho(\lambda)
{ m\over{\lambda^2+m^2}}
\eqno(3.1)
$$
where $\rho(\lambda)\geq 0$ is the eigenvalue density of the Dirac operator.
We define the two susceptibilities

$$
\chi_\sigma=\int_x <\bar\psi\psi(x)\bar\psi\psi(0)>,\,\,\,\,\,\,
\chi_\pi=\int_x <\bar\psi i\gamma_5 \psi(x)\bar\psi i\gamma_5\psi(0)>
\eqno(3.2)
$$
which are the zero-momentum projections of the
scalar and pseudoscalar propagators. Again, we emphasize that the scalar
correlation function is understood as the connected one.
These two susceptibilities are related to the order parameter
via: $\chi_\sigma=\partial <\bar\psi\psi>/\partial m$ and the Ward identity
$\chi_\pi=<\bar\psi\psi> /m$. (We will
derive this relationship later in this section.)
With this in mind, it is straightforward to
show that these susceptibilities can be rewritten in terms of the spectral
function as

$$
\chi_\sigma=\chi_\pi-\int_{-\infty}^{+\infty}d\lambda\rho(\lambda)
{ {2m^2}\over{(\lambda^2+m^2)^2} }
\eqno(3.3)
$$
The integral on the right hand side is positive leading to the following
inequality $\chi_\sigma \leq \chi_\pi$.
Since the mass, defined as in eq.(2.4),
is related to the susceptibility via $M^2=Z\chi^{-1}$,
and since scalars and pseudoscalars renormalize in the same way, we
obtain the following inequality
(level ordering): $M_\pi^2\leq M_\sigma^2$.

{}From eqs.(3.1) and (3.3) it is apparent that both, the order parameter and
the
$\sigma-\pi$
splitting disappear in the chiral limit if the
zero-mode is absent from the spectrum i.e. when $\rho(0)=0$. Conversely,
the presence of the zero-mode
induces both a nonzero order parameter,
$<\bar\psi\psi>=\pi\rho(0)\not=0$,  and $\sigma-\pi$ splitting.
The same mechanism that leads to a nonvanishing value of the
order parameter is responsible for lifting the degeneracy within the
parity doublets.

The starting point in our analysis
is the scaling form of the equation of state [8,9].
In the case of continuous symmetry
the familiar form $h=M^\delta f(t/M^{1/\beta})$ becomes

$$
h_a=M_a M^{\delta-1} f(t/M^{1/\beta})
\eqno(3.4)
$$
where, for simplicity, we denote the vector order
parameter and external field by $M_a$ and $h_a$ respectively, and the modulus
of the order parameter as $M=\sqrt{M_a M_a}$. Our convention is
that $t>0$ corresponds to the broken phase.Eq.(3.4)
is the gap equation for the order parameter.
In the limit when $h\to 0$ we have

$$
0=M^\delta f(x)
\eqno(3.5)
$$
which has a nontrivial solution $M\not= 0$ only if the function $f(x)$ has
a positive zero $f(x_0)=0$. The spontaneous magnetization is given
by $M=(t/x_0)^\beta$. This determines only the modulus of the order parameter
(vacuum degeneracy), whereas the orientation and, thus, the
particular Hilbert space is fixed
by the direction of the
external field. The function $\tilde f(x/x_0)=f(x)/f(0)$ is the same
for any theory within a given universality class. At the critical point $x=0$
the response of the system is singular and is given by $h=M^\delta f(0)>0$.
In order to have no spontaneous magnetization in the symmetric phase
$f(x)$ should be positive on the negative x-axis.

The response to an external field $h_a$ is given by the inverse susceptibility
$(\chi^{-1})_{ab}=\partial h_a/\partial M_b$
which can be obtained from the equation of state

$$
(\chi^{-1})_{ab}=\delta_{ab}M^{\delta-1} f(x)
+(\delta-1) {{M_aM_b}\over{M^2}}   M^{\delta-1} f(x)
-{x\over\beta} {{M_aM_b}\over{M^2}}      M^{\delta-1} f'(x)
\eqno(3.6)
$$
where $x=t/M^{1/\beta}$. Eq.(3.6) can be rearranged into

$$
(\chi^{-1})_{ab}=\biggl(\delta_{ab}-{{M_aM_b}\over{M^2}}\biggr)
M^{\delta-1} f(x)
+{{M_aM_b}\over{M^2}} M^{\delta-1} \biggl(\delta f(x)
-{x\over\beta}f'(x)\biggr)
\eqno(3.7)
$$
in order to separate the susceptibility into
transverse and longitudinal parts

$$\chi^{-1}_T= M^{\delta -1} f(x), \,\,\,\,\,\,\,\,
\chi^{-1}_L =M^{\delta-1} \biggl(\delta f(x)
-{x\over\beta}f'(x)\biggr)
\eqno(3.8)
$$
The expression for the transverse susceptibility is
the Ward identity which can be rewritten, after using the EOS, as

$$h=M\chi_T^{-1}
\eqno(3.9)$$
In the broken phase where
$M\not= 0$, the divergence of the transverse susceptibility in the $h\to 0$
limit signals the appearance of massless modes (Goldstone bosons).
At the critical point $(x=0)$, eqs.(3.8) imply
that

$$
{ {\chi^{-1}_L}\over{\chi^{-1}_T} }=\delta
\eqno(3.10)
$$
This defines the exponent $\delta$ as a
measure of the relative strength of longitudinal and transverse
responses of the system at the critical point. Eq.(3.10) can
also be derived by
noting that, at the critical point, $M\sim h^{1/\delta}$, and
using $\chi_L=\partial M/\partial h$ and $\chi_T=M/h$.

When the symmetry in question is
chiral symmetry, eqs.(3.8)
can be combined into a statement about the meson masses.
The external field is replaced by the bare fermion mass $m$ and
the order parameter, $M_a$,  has components $(<\bar\psi\psi>,
<\bar\psi i\gamma_5 T^a \psi>)$ for the flavor group whose generators are
$T^a$. ( The correct mapping to ensure dimensional
consistency is $<\bar\psi\psi> \to \Lambda^2M$ and $h  \to m\Lambda^2$.
In the following we will work with dimensionless quantities, so we are allowed
to take $\Lambda = 1$ from here on).

We introduce the bare mass in the standard way through $m\bar\psi\psi$,
so that the ground state is
parity invariant. Then, condensation occurs in the scalar
channel and the only nonvanishing component of the order parameter is
$<\bar\psi\psi>$. The EOS, eq. (3.4), then reads
$$m = <\bar\psi\psi>^\delta f (t/<\bar\psi\psi> ^{1/\beta})\eqno(3.11)$$
The sigma and the pion are longitudinal and
transverse modes, respectively. The Ward identity, eq.(3.9), now becomes
$$
<\bar\psi\psi>=m\int_x <\bar\psi i\gamma_5 \psi (x)
\bar\psi i\gamma_5 \psi (0)>
\eqno(3.12)
$$
The masses are related to the corresponding
susceptibilities via $M_\pi^2 =Z_\pi \chi_\pi^{-1}$,
$M_\sigma^2 =Z_\sigma \chi_\sigma^{-1}$.
Because of chiral symmetry, the wavefunction renormalization
constants for the two modes are equal and the mass ratio (squared) is the
susceptibility ratio i.e.

$$R_\pi=M_\pi^2/M_\sigma^2
=\chi^{-1}_T/\chi^{-1}_L
\eqno(3.13)
$$
Thus, eqs.(3.8) implies

$$
{1\over{R_\pi(t,m)}}=\delta - {x\over\beta}{ {f'(x)}\over{f(x)} },
\,\,\,\,\,\,\,\,
R_\pi(0,m)={1\over\delta}\,\,\,\,\,\,
\eqno(3.14)
$$
Eq.(3.14) is a stronger statement than the Goldstone theorem
eqs.(3.9) or (3.12). Not only does it contain the
above Ward identity, but it determines the splitting within the parity
multiplet away from the chiral limit in both phases.
Before we discuss the above result, we recall
the EOS for the masses,

$$
M_\sigma(t,m)=t^\nu g_\sigma\bigl({m\over{t^\Delta}}\bigr),\,\,\,\,\,\,\,\,
M_\pi(t,m)=t^\nu g_\pi\bigl({m\over{t^\Delta}}\bigr)
\eqno(3.15)
$$
While the equation for $M_\sigma$ follows from hyperscaling, the same
reasoning does not lead to the analogous expression
for $M_\pi$. Its form is fixed, however, by eq.(3.13)
which states that the ratio $R_\pi$ is a function of $x$ and, thus, of
$m/t^\Delta$, implying eq.(3.15).
The functions $g(y)$ are universal up to a multiplicative constant.
Clearly, the Goldstone nature of the pion imposes a different boundary
condition on $g_\pi(y)$ requiring $g_\pi(y)\sim y^{1/2}$ near the origin.
The argument $m/t^\Delta$ is understood
in the sense $t^\Delta \to sgn(t) |t|^\Delta$
so that it changes sign as one goes through the critical point.
This insures single-valuedness of $g(y)$ and the possibility
of global inversion. Taking the mass ratio
eliminates non-universal multiplicative factors as well as
the exponent $\nu$ making it a universal function of $y=m/t^\Delta$ only

$$
R_\pi(t,m)={ {M_\pi^2}\over{M_\sigma^2} }=
G\bigl({m\over{t^\Delta}}\bigr)
\eqno(3.16)
$$
In this equation, we have the low energy observable on one side and
the bare parameters on the other. From eq.(3.14) it follows that

$$
R_\pi(t=0,m)=G(\infty)={1\over\delta}
\eqno(3.17)
$$

There are two limiting values of $R_\pi$ that are
fixed by chiral symmetry.
In the broken phase $(t>0$) pions  are Goldstone bosons, so
$R_\pi(t>0,m=0)=0$. In the symmetric phase the sigma and pion degeneracy
implies $R_\pi(t<0,m=0)=1$.
The ratio can never exceed unity since
Euclidean propagators satisfy $|D_\pi(x)|\ge |D_\sigma(x)|$ which, in the
large $|x|$ limit gives $R_\pi\le 1$. As the bare mass
increases, the ratio
approaches the same value regardless of which phase it originates from. This
is so simply because, as one increases the amount of explicit symmetry
breaking,
the dynamics becomes insensitive to the type of symmetry realization
in the vacuum. Thus,
if we fix $t$ and plot the curve $R_\pi(t,m)$ versus $m$
(Fig.1), in the broken
phase $R_\pi=R_>$ will increase from zero
while in the symmetric phase $R_\pi=R_<$ will
decrease with $m$ starting from $R_\pi=1$.
Both families of curves approach $R_c=1/\delta$
asymptotically from above (symmetric) and below (broken).
Because of the scaling
form of eq.(3.16),
$R_\pi(t,m)=G(m/t^\Delta)$, it is clear that $R_c=G(\infty)$ is
independent of $m$. So, we have traded small $t$ for large $m$.
Since $0\leq R_\pi \leq 1$, it follows that $R_>$ and
$R_<$ curves have slopes with
opposite signs. Thus, all the curves
in the broken phase are monotonically increasing and
lie below $R=R_c$ i.e. $R_>(m)<R_c$ and $R'_>(m)>0$ for all $m$.
Similarly, in the symmetric phase $R_<(m)>R_c$ and $R'_<(m)<0$.
Therefore, the following inequality holds

$$
R_>(m)< R_c < R_<(m)
\eqno(3.18)
$$
which is saturated asymptotically for large values of $m$.
Since $R_c =1/\delta$, eq.(3.18) gives a bound on the value of $\delta$

$$
{1\over{R_<(m)}} <\delta<{1\over{ R_>(m)}}
\eqno(3.19)
$$
Any curve $R_\pi (m)$ in the
broken phase produces an upper bound on $\delta$ etc..
This bound improves as the ratio $m/t^\Delta$ becomes larger.
The analysis of the mass ratio $R_\pi$ is consistent with the function
$f(x)$ being positive semi-definite with only one zero at $x_0$
(Fig.2). A possible
change in its monotonicity would imply the vanishing of its first derivative
for at least one non-zero
value of $x$. This would be in conflict with the physical
behavior of $R_\pi$.

There are several reasons why this result is useful. Its application to
data analysis produced by lattice simulations is obvious.
Recent applications of this method to strongly
coupled $QED$ have been very successful [14], giving more accurate
estimates of the critical coupling and the exponent $\delta$
than those obtained with conventional methods
on data samples of comparable statistics.

The reason for this can be easily understood. Instead of dealing with
extrapolated data, as normally done when studying chiral symmetry breaking
through $<\bar\psi\psi>$, here we determine both, the critical coupling and
critical exponent $\delta$ from the raw data. Furthermore, the scaling form
for the mass ratio is more accurate for larger values of $m/t^\Delta$. This
means that, instead of simulating at the critical point and looking at the
$m\to 0$ limit, we can work away from it and use large masses without
losing accuracy.
The key point, from the point of view of
numerical simulations, is the $m$-independence of $R_\pi (t, m)$ at the
critical point, eq (3.17). This can exploited by plotting $R_\pi (t,m)$
as a function of $\beta$ for different $m$ values. The spectral method
can be used to get very accurate plots, which cross at the critical point,
thus determining $\beta_c$. This is reminiscent of the techniques
of Finite Size Scaling when applied to the Binder
cumulant, as noticed also by Boyd et al.
in ref. [15] where a similar method  has been  proposed and
successfully applied
to the $QCD$ chiral transition in the strongly coupled phase.

\vskip10truemm

\noindent{4. PCAC, HYPERSCALING AND TRIVIALITY}
\vskip5truemm

Eq.(3.14) is a differential equation that determines
the function $f(x)$ in terms of other universal
quantities like critical exponents and mass ratios. Its solution requires
one boundary condition which can be fixed at an arbitrary value of $x$.
The formal solution for $f(x)$ is given
by the form

$$
f(x)=f(0)\exp\biggl(\beta\int_0^x {{dy}\over y} \bigl(\delta
-{1\over{R_\pi}}\bigr)\biggr)
\eqno(4.1)
$$
which shows that the ratio $f(x)/f(0)$ is universal.
In the symmetric phase,
$x<0$, $R_\pi \to 1$ in the chiral limit ($x\to -\infty$).
Therefore, in this limit
eq.(3.14) gives

$$
\gamma=\beta(\delta-1)=\lim_{x\to -\infty} { {xf'(x)}\over{f(x)} }
\eqno(4.2)
$$
where we used the scaling relation between the critical exponents:
$\gamma=\beta(\delta-1)$. The above equation implies that, for $x\to-\infty$,
$f(x)\approx C (-x)^\gamma$. In general, the expansion of $f(x)$ in the
symmetric phase has the form $f(x)=\sum_n a_n (-x)^{\gamma-2n\beta}$ [16].

Another constraint on the behavior of $f(x)$ is PCAC. It refers to the
chiral limit in the broken phase. Eq.(3.14) can be rewritten as

$$
R_\pi = {  {\beta f(x)}\over{\delta\beta f(x)-xf'(x)}  }
\eqno(4.3)
$$
which implies the vanishing of the pion mass in the chiral limit $(x\to x_0)$.
Now, we show how PCAC
forces $x_0$ to be a first order zero. Assume that around $x_0$,
$f(x)$ vanishes as

$$
f(x)\approx a(x_0-x)^\rho
\eqno(4.4)
$$
In the chiral limit (in the broken phase) eq.(4.3) becomes

$$
R_\pi = {  {\beta(x_0-x)}\over{(\delta\beta-\rho)(x_0-x)+\rho x_0}  }
\eqno(4.5)
$$
so $M_\pi^2$ vanishes linearly.
For $x\approx x_0$ the equation of state is

$$
{m\over{<\bar\psi\psi>^\delta}}=f(x)\approx a(x_0-x)^\rho
\eqno(4.6)
$$
Clearly, PCAC, which requires $M_\pi^2 \sim m$,
forces $\rho=1$. Substituting eq.(4.6)
into the expression for $R_\pi$ (eq.(4.3)) gives

$$
R_\pi = \beta{m\over{(\delta\beta-1)m+ax_0<\bar\psi\psi>^\delta}}
\eqno(4.7)
$$
As $m\to 0$ the leading contribution is

$$
R_\pi\approx{\beta\over{ax_0}} {m\over{<\bar\psi\psi>^\delta}}
\eqno(4.8)
$$
When compared with the PCAC relation,
$M_\pi^2=2m<\bar\psi\psi>/f_\pi^2$, eq.(4.8) gives

$$
{{f_\pi^2}\over{M_\sigma^2}}={{2x_0}\over\beta}
{{a<\bar\psi\psi>^{\delta+1}}\over{M_\sigma^4}}
\eqno(4.9)
$$
This relates the pion decay constant to the physical scale. We can use
the definition of the critical exponents to study the behavior of $f_\pi$
in the scaling region e.g. $\beta(\delta+1)=2\Delta-\gamma$.
However, using the EOS for $<\bar\psi\psi>$ (eq.(3.11)) and the EOS for
$M_\sigma$ (eq.(3.15))
the above relation can be written as

$$
{{f_\pi^2}\over{M_\sigma^2}}= M_\sigma^{(2\Delta-\gamma-4\nu)/\nu}
K(m/t^\Delta)
\eqno(4.10)
$$
with $K(y)$ being a universal function.
The exponent is the same one that appears in the expression for the
renormalized
coupling (eq.(2.9)).
Clearly, since the above ratio is dimensionless, hyperscaling
implies the vanishing of the exponent.
Conversely, its violation would imply that the above
ratio is cutoff dependent.
In theories with Yukawa couplings, triviality is intimately related to
the issue of compositeness of the scalars. For a nontrivial continuum limit
to exist, it is necessary for fermions to exchange composite scalars. This is
contained in the Goldberger-Treiman relation $g=M_F/f_\pi$, which, after
recognizing that the pion radius scales as $r_\pi\sim 1/f_\pi$ implies that
the coupling vanishes in the limit where pions are pointlike
$(g\sim M_F r_\pi)$. In that sense,
compositeness means that the pion decay constant has to scale as the physical
mass scale. Pointlike structure, on the other hand, implies that $f_\pi$
diverges in physical units. Although Baker's inequality is rigorously
proven only for ferromagnetic systems [10]
%[B,S]
, it is interesting to note that it
is valid in the above equation in all known models where $f_\pi/M_\sigma$
either diverges or approaches a constant
depending on whether pions are pointlike or
not. The inequality $\Delta\leq \gamma+4\nu$ is in agreement with it.
\vskip1truecm

\noindent
5. PION MASS, ORDER PARAMETER AND ANOMALOUS DIMENSIONS
\vskip5truemm

Nonvanishing anomalous dimensions and compositeness are
intimately related. In addition, the
compositeness condition itself is tied to the existence of a
fixed point of the underlying theory. It is
reflected in the vanishing of the
wave function renormalization constant associated with the composite
degrees of freedom. Let us explain this in some detail. Consider a
general field theory where a bound state $|B>$, with binding
energy $E_B <0$, appears. From general considerations [17]
%[W]
, the wave function
renormalization constant $Z$ can be expressed as

$$
Z=\sum_{b,E} |<b,E|B>|^2
\eqno(5.1)
$$
where $|b,E>$ is the bare (elementary particle) state with energy $E$.
Using standard
techniques from quantum mechanics, it can be shown that $Z$ satisfies the
following equation [17]
%[W]

$$
1-Z=\int_0^\infty dE { {G^2(E)}\over{(E+|E_B|)^2} }
\eqno(5.2)
$$
with $G^2(E)>0$ being the total decay rate of the state $|B>$,
$G^2(E)=2\pi \sum_b |<b,E|B>|^2$. As such $G(E)$ is proportional to an
effective coupling constant. The compositeness condition, $Z=0$, implies
two things: 1) The composite state has no projection in the space of
bare states (eq.(5.1)) i.e. $<b|B>=0$ for any $|b>$, and 2)
It is a sum rule that places an upper bound on
the effective coupling, $G^2(E)$, (eq.(5.2)) and can be
interpreted as a fixed point condition.

As the fixed point is approached, the order parameter, wave function
renormalization constant, $Z$, and all the masses vanish. If there is a single
correlation length, the vanishing of $Z$ is determined by the anomalous
dimension, $\eta$, via $Z\sim \xi^{-\eta}$. Also, the order parameter,
$<\phi>$ scales as $<\phi> \sim \xi^{-d_\phi}$, where
$d_\phi =(d-2+\eta)/2$ is the scaling dimension of the field $\phi$.
Combined with the standard scaling laws in the symmetric limit,
$<\phi> \sim t^\beta$ and $\xi\sim t^{-\nu}$, this leads to the following
scaling relation between the exponents

$$
{\beta\over\nu}={ 1\over 2}(d-2+\eta)
\eqno(5.3)
$$

The appearance of the anomalous dimensions guarantees the compositeness
of the degrees of freedom involved, a
necessary condition to produce a nontrivial continuum limit.
Consider the pair of equations used in
previous sections

$$
m=<\bar\psi\psi>^\delta f(x), \,\,\,\,\,\,
\chi^{-1}_\pi=<\bar\psi\psi>^{\delta-1}f(x)
\eqno(5.4)
$$

We shall see how eqs.(5.4) restricts
the dependence of $M_\pi$ on the order parameter in the
chiral
limit and show
that this behavior is universal and is determined by the magnitude of the
anomalous dimension $\eta$. As such, it contains information about the
continuum limit of the theory and is capable of distinguishing between
mean field and non-mean field behavior.
This being the case, perhaps, the best way to start
is to recall this dependence in the $\sigma$-model in four dimensions.
The masses and
EOS are given by

$$
M_\pi^2=-t+{\lambda\over 6}v^2,\,\,\,\,\,\,
M_\sigma^2=-t+{\lambda\over 2}v^2,\,\,\,\,\,\,
h=v\bigl(-t+{\lambda\over 6}v^2\bigr)
\eqno(5.5)
$$
where $v=<\sigma>$ is the order parameter. The universal function in this case
is a straight line $f(x)=-x+\lambda/6$.
In the chiral limit the order parameter is obtained from $f(x_0)=0$ giving
$v_0^2=6t/\lambda $.
Since this is a mean field model, there is no
wavefunction renormalization and $M_\pi^2=\chi_\pi^{-1}$. Eq.(5.5) thus,
reproduces the Ward Identity and EOS
of eq.(5.4). For a fixed value of $t$, $M_\pi^2$ is a linear function
of $v^2$ with different intersections depending on the phase .
The expression for the pion mass can be rewritten as
$M_\pi^2=(\lambda/6)(v^2-v_0^2)$. Thus,
as discussed in the previous section,
the pion mass vanishes linearly  as the
chiral limit is approached . In the
symmetric phase the $v\to 0$ limit results in a nonvanishing pion mass .
Again, the chiral limit is approached linearly.
The reason for such a simple behavior is the absence of
wavefunction renormalization (vanishing anomalous dimensions).
It is easily demonstrated that, as long as the identification $M_\pi^2=
\chi_\pi^{-1}$ can be made, eqs.(5.4) insures the linear dependence between
$M_\pi^2$ and $<\bar\psi\psi>^2$. To show this, we recall the expansion
of the universal function $f(x)$ around the chiral limit in the symmetric
phase:
$f(x)\sim |x|^\gamma$, which leads, in the limit $<\bar\psi\psi>\to 0$,
to $M_\pi^2 \sim |t|^\gamma$. The general dependence of the pion mass on
the order parameter for a mean field theory is summarized in Fig.3.

Since, for nonvanishing anomalous dimension $\eta>0$, the
wavefunction renormalization constant scales as well,
it will give some curvature to $M_\pi^2$ near the
origin both in the symmetric phase and at the critical point
(Fig.4). This curvature
will be a signal of nontrivial behavior of the theory
and is related to the compositeness of the pions.
This curvature is 1) determined by the magnitude
of the anomalous dimensions and 2) is opposite from that induced by finite
size effects. In this way, the above plot contains information about both
the thermodynamic limit and nontriviality of the theory.

In general, the pion mass is related to the susceptibility as
$M_\pi^2=Z\chi^{-1}_\pi$.
The wavefunction renormalization constant scales
as $Z\sim \xi^{-\eta}$ and, since the order parameter scales as
$<\bar\psi\psi>\sim \xi^{\beta/\nu}$, we have
$Z\sim <\bar\psi\psi>^{\nu\eta/\beta}$.
At the critical point, $t=0$, $\chi^{-1}_\pi\sim <\bar\psi\psi>^{\delta-1}$.
Thus,

$$
M_\pi^2 \sim <\bar\psi\psi>^{\delta-1+\eta\nu/\beta}
\eqno(5.6)
$$
The exponent can be expressed in a slightly different form using the scaling
relations between the critical exponents: $\beta(\delta-1)=\gamma$ and
$\gamma=\nu(2-\eta)$. This leads to

$$
\delta-1+{ {\eta\nu}\over\beta}={1\over\beta}(\gamma+\eta\nu)
={{2\nu}\over\beta}=2\biggl(1- { {d-4+\eta}\over{d-2+\eta} }\biggr)
\eqno(5.7)
$$
In four dimensions, the expression in parenthesis is $1/(1+\eta/2)<1$.
Thus, at the critical point we have

$$
M_\pi^2\sim (<\bar\psi\psi>^2)^{\nu/\beta},\,\,\,\,\,\, (t=0)
\eqno(5.8)
$$
The important point is that the slope of the curve is infinite near the origin
because of the anomalous dimensions: $\beta/\nu=1+\eta/2$ in four dimensions.

In the symmetric phase, we use the asymptotic expansion of the universal
function $f(x)$ for $x\to -\infty$ [16]:

$$
f(x)=\sum_{n=0}^{\infty} a_n |x|^{\gamma-2n\beta}
\approx a_0|x|^\gamma +a_1|x|^{\gamma-2\beta}
\eqno(5.9)
$$
The pion susceptibility and its mass are determined from this expression giving

$$
M_\pi^2\sim <\bar\psi\psi>^{\delta-1+\eta\nu/\beta}
\biggl( a_0{{|t|^\gamma}\over{<\bar\psi\psi>^{\gamma/\beta}}}
+a_1{{|t|^{\gamma-2\beta}}\over{<\bar\psi\psi>^{(\gamma-2\beta)/\beta}}}\biggr)
\eqno(5.10)
$$
The first term gives the leading contribution near the origin. To obtain its
magnitude, we use once again $\beta(\delta-1)=\gamma$. In
that case the leading term gives

$$
M_\pi^2\sim (<\bar\psi\psi>^2)^{\eta\nu/2\beta},\,\,\,\,\,\, (t<0)
\eqno(5.11)
$$
{}From the scaling relations it follows that the exponent in eq.(5.11)
is less than 1:

$$
{{\eta\nu}\over{2\beta}}= 1-{{d-2}\over{d-2+\eta}}<1
\eqno(5.12)
$$
Clearly, for vanishing $\eta$, the leading term is constant and the second term
in eq.(5.10) determines the curvature. The exponents in that case
are given by mean field theory, so $1/\beta=2$.
Therefore, $M_\pi^2$ approaches some finite value linearly. The curvature,
and the infinite slope come from the anomalous dimensions.

Finally, in the broken phase, $t>0$, the order parameter
is nonvanishing in the $m\to 0$ limit and the function $f(x)$ vanishes for
$x=x_0$. PCAC constrains this to be a first order zero i.e.
$f(x)\approx a(x_0 - x)$. Therefore, in the
chiral limit we have

$$
M_\pi^2=Z\chi^{-1}_\pi \sim <\bar\psi\psi>^{\delta-1+\eta\nu/\beta}a(x_0 -x)
\sim (<\bar\psi\psi>-<\bar\psi\psi>_0), \,\,\,\,\,\, (t>0)
\eqno(5.13)
$$
and the pion mass vanishes linearly.

A remark about the apparent vanishing of the pion
mass in the symmetric phase (in the chiral
limit) should be made. At first glance
this result is surprising, because
symmetry considerations alone do not force its vanishing.
In mean field theory where
$M_\pi^2=\chi_\pi^{-1}$ everything is
canonical and equation (5.10 )
tells us that $M_\pi^2$ is finite in the symmetric phase even for $m=0$:

$$
\chi^{-1}_\pi
\sim <\bar\psi\psi>^{\delta-1} (|t|/<\bar\psi\psi>^{1/\beta})^\gamma
\sim |t|^\gamma
\eqno(5.14)
$$
The vanishing comes from the anomalous
dimensions i.e. wavefunction renormalization! Let us analyze this point
in more detail. The correlation length in Euclidean theory is

$$
2d\xi^2={  {\int_x |x|^2 D(x)}\over{\int_x  D(x)}  }
\eqno(5.15a)
$$
or, in momentum space

$$
2d\xi^2=\biggl( {1\over{D^{-1}(k^2)}}
{ {dD^{-1}(k^2)}\over{dk^2} }\biggr)_{k^2=0}
\eqno(5.15b)
$$
This is a standard way of extracting the
infra-red piece of the propagator that
dominates at large separations. If we denote the self-energy corrections
to the meson propagator as $\Pi(k^2)$, the full propagator in momentum
space reads

$$
D^{-1}(k^2)=k^2+M_0^2+\Pi(k^2)=k^2 +M_0^2+\Pi(0)+k^2\Pi'(0) +\tilde\Pi(k^2)
\eqno(5.16)
$$
This decomposition is made so that $\tilde\Pi(k^2)=O(k^4)$ and the first
two terms dominate the infra-red region. After identifying the $k^2$ term
in eq.(5.15b) as the inverse wavefunction renormalization constant,
$Z^{-1}=1+\Pi'(0)$ and inverse susceptibility as
$\chi^{-1}=M_0^2+\Pi(0)$, the meson propagator can be written as

$$
D^{-1}(k^2)=Z^{-1}(k^2 +Z\chi^{-1}+\tilde\Pi_R(k^2))\approx
Z^{-1}(k^2 +M^2+O(k^4))
\eqno(5.17)
$$
Thus, at low momenta, the propagator resembles that of a meson with
mass $M^2=Z\chi^{-1}$. This is the definition of the mass as given by eqs.
(5.15).
In this case analytic continuation to Minkowski space is standard
and the pole structure is recovered. As is obvious from the decomposition
in eq.(5.16), such manipulations assume a particular
analytic structure of the composite propagators which is guaranteed
only if the fermions are massive. Additional nonanalyticities
in the form of branch cuts appear
when fermions are massless as occurs in the chiral limit in the
symmetric phase. The anomalous scaling of the scalar propagator is a
consequence of these nonanalyticities -- the $k^2$ term is absent. Rather,
the leading low-momentum behavior is given by

$$
D^{-1}(k^2)=k^{2-\eta}+C
\eqno(5.18)
$$
The derivative diverges for $k^2=0$ giving
zero mass as defined by eqs.(5.15), although there might be a pole in
Minkowski space. Such
a propagator does not have an exponential, but a power law decay.
So, there are long-range correlations (``massless modes''),
but the theory is not scale invariant.
(Clearly, the case $\eta=0$ is well behaved. In the
broken phase there are no infra-red "problems" of this sort
because fermions are massive and eq.(5.15) is the appropriate definition of
the correlation length.) Some attempts to confront this
problem in a lattice
study of three-dimensional Gross-Neveu model have been reported in [4].
%[HKK].

When studying the theory in a finite volume, as is always done in lattice
simulations, one has to insure that results are not obscured by
finite size effects. Since different
quantities have different sensitivity to finite
size effects, the effective thermodynamic limit, therefore,
must be monitored carefully through the consistency of certain
relations. One set of such relations consists of
the Ward Identity and the Equation of State.
Being the lightest particle in the spectrum, the pion is most sensitive to
finite size effects. On a finite lattice,
the pion mass would tend to a higher value
than in the thermodynamic limit. This would result in a change of sign in the
curvature. In that sense, the plot $M_\pi^2$ vs $<\bar\psi\psi>^2$ is
also suited for controlling finite size
effects in theories with nontrivial fixed points.
\vskip10truemm

\noindent{6. HEAVY QUARK LIMIT AND EXPONENT $\delta$}
\vskip5truemm

We have seen that $R_\pi = G(m/t^\Delta)$
with $G(\infty)=1/\delta$. This means that, in
the large-$m$ limit, $R_\pi\to 1/\delta$ (not
necessarily 1). At first sight this looks puzzling since one expects that
the nonrelativistic limit can be taken for very large constituent
masses so that
any meson mass should approach twice the constituent mass i.e. $M_{meson}
\approx 2m$ which would necessarily imply $R_\pi\approx 1$.
We should clarify what is meant by a large fermion mass: the
bare mass is always much smaller than the cutoff and large $m$
means large compared to a typical mass scale, for example $M_\sigma$
in the chiral limit. In this way we are sure to stay in the
scaling region where universality arguments hold and the RG trajectories are
uniquely defined.

In theories
that are not asymptotically free
%, the heavy quark limit does not really
%exist in the scaling region.
expansions in powers of $p/m$ are poorly behaved
because the force between constituent quarks is strong at
short distances forcing a large kinetic energy due to
uncertainty relations. Thus, whatever the quark mass, it is never large
compared to a typical kinetic energy. The composites in the scaling region
are always relativistic. We note also  that the bound states
in the two phases are quite different
from each other. In the symmetric phase,
an increase in the bare mass is compensated by a decrease in the coupling
in order to keep the ratio unchanged.
This is expected
because the zero-point energy
is reduced by increasing the mass and
less attraction is needed to produce the same effect.
In the broken phase, the opposite happens.

In theories like QCD this is not so because the
force between the quarks is weak at short distances and the typical momenta
could be small compared to the bare mass. Thus, in principle, one can still
be in the scaling region and have a heavy quark limit. The most direct way
of understanding
the differences between asymptotically free and non-asymptotically
free theories is to note that in the scaling region of the former $|t|\ll 1$
means weak coupling $(g_c=0, t=g^2)$, whereas for the latter $|t|\ll 1$ means
strong coupling.

We should add that $\delta=1$ has different physical origins
in asymptotically free and non-asymptotically free
theories. Since the exponent $\delta$ gives the response of the system at the
critical coupling: $m=<\bar\psi\psi>^\delta |_{t=0}$,
in asymptotically free theories the
system responds as a free theory, so $\delta=1$.
In that context, $R_\pi\approx 1$ is a reflection
of the fact that there is very weak binding which could not
possibly distinguish between the scalars and pseudoscalars.
In strongly coupled non-asymptotically free
theories, tightly bound relativistic composites are
formed in the scaling region. Their presence at high energies is the main
difference relative to QCD-like theories where the only relevant
degrees of freedom in the scaling region
are quarks and gluons
%[]
and where
binding occurs in the infra-red regime.
Because of the non-asymptotically free nature of
the couplings, the ultra-violet asymptotics of
the scalar correlation functions
is not canonical -- the theory has anomalous dimensions. In terms of the
anomalous dimensions the exponent $\delta$ in four dimensions is

$$
\delta={{6-\eta}\over{2+\eta}}
\eqno(6.1)
$$
Here, $\delta=1$ is a consequence of the large anomalous dimension, $\eta=2$.
The meaning of this particular limit can be best understood if we write down
the first two terms in the operator product expansion of the fermion
propagator [18]
%[CG]

$$
S(p)\approx  {A\over{p \!\!\! /}}+
{{Bm}\over{p^{2+\eta/2}}} +{{C<\bar\psi\psi>}\over {p^{4-\eta/2}}}+...
\eqno(6.2)
$$
As $\eta\to 2$ i.e. $\delta\to 1$, the system reacts to
the bare mass the same way it would react to a change in the
dynamical mass of its constituents.
Thus, the persistence of $R_\pi=1$ away from the chiral limit
in this case means that
the system can not distinguish between bare and dynamical masses.

As a final remark, we use once again eq.(3.14)
to argue the absence of the dilaton
and scale invariance in theories with spontaneously broken chiral symmetry.
Unlike previous treatments [19]
%[M,BLL]
our argument will require no knowledge
of the composite propagators and is thus independent of the approximation
scheme. Instead it follows from the scaling form of the EOS only.
The idea of a dilaton as a Goldstone boson of spontaneously broken scale
symmetry
was introduced some time ago in ref.[20].
If the theory is scale invariant
and if it breaks chiral symmetry spontaneously, it, at the same time, generates
a scale, the fermion mass, and is no longer scale invariant. Since scale
invariance is thus
broken "spontaneously",
one naively expects that there should be a corresponding Goldstone boson, a
massless scalar, in the spectrum.

We use the scaling relation $\beta=\gamma /(\delta-1)$ to rewrite
eq.(3.14) as

$$
R_\pi={1\over {\delta - x(\delta-1)/\gamma \bigl(f'(x)/f(x)\bigr)}}
\eqno(6.3)
$$
In general, it is clear that there has to be splitting between the $\sigma$
and $\pi$ i.e. there can be no massless scalar in the theory as long as
chiral symmetry is realized in the
Nambu Goldstone manner. An exception might only be
the theory with $\delta=1$. In that case $R_\pi=1$ for any
value of the bare parameters -- not only are the parity partners degenerate,
but
they respond to symmetry breaking in the same way.
This would be a very unusual realization of chiral symmetry: $\sigma$ and $\pi$
are indistinguishable for any finite $m$. In QCD, where $\delta=1$, this
situation is avoided by explicit violations of scaling due to quantum
corrections. These scaling violations are the sole source of interaction
and of the $\sigma-\pi$ splitting.
In other cases e.g. strongly coupled QED with $\delta=1$, the
$\sigma-\pi$ degeneracy in the broken phase could not be
reconciled with PCAC which requires $R_\pi \sim m$. Thus, if the
current algebra relations are to be realized, scale invariance must be
violated. The entire content of PCAC is contained in these scaling violations.
Consequently, the
mass of the $\sigma$ comes solely from the scaling violation.
\vskip10truemm

\noindent{7. TWO EXAMPLES (SCALING PLOT) }
\vskip5truemm

To illustrate and complement the previous discussion, we analyse two simple
examples: the four-fermi and linear $\sigma$ models.
In addition to demonstrating realizations of the general ideas in these
two models, we also discuss the mutual
dependence of the mass ratios , or scaling plots, (as introduced
in lattice  spectroscopy in ref. [21]),
and the heavy quark limit in these models. The scaling plot will
prove especially suitable to argue the equivalence of the two models.
(For notations and details of computations related to this
Section see [13].)

First, we start with the four-fermi model.
In the leading order in $1/N$ there
is no distinction (on the technical level) between discrete and continuous
chiral symmetry. We consider the continuous $U(1)\times U(1)$ model.
For $2<d<4$ these models are renormalizable
and nontrivial. To leading order, the critical
exponents are $\beta=\nu = 1/(d-2), \delta=d-1,\eta=4-d$.
The gap equation for the
fermion self-energy $\Sigma$ is given by the tadpole contribution

$$
\Sigma = m+4g^2<\bar\psi\psi>
\eqno(7.1)
$$
or, explicitly, in terms of $\Sigma$,
$$
cg^2\Sigma^{2-\epsilon} = t+{m\over\Sigma}
\eqno(7.2)
$$
where $\epsilon=4-d$ (not necessarily small, $0<\epsilon <2$) and
$c=4b/(2-\epsilon)$, with $b=2\Gamma (\epsilon /2)/(4\pi)^{d/2}$.
We will express all the quantities in units of the momentum
cutoff $\Lambda=1$. The masses
of the composites $M_\sigma,M_\pi$ are given by the one-loop diagrams [13]

$$
M_\pi^2 = {m\over\Sigma} Z, \,\,\,\,\,\,
M_\sigma^2 =M_\pi^2 +4\Sigma^2
\eqno(7.3)
$$
with the wave function renormalization constant $Z=\Sigma^\epsilon/bg^2$.

We first discuss the dependence of $M_\pi^2$ on the order parameter.
In this case, the expectation value of the scalar field can be used
to make contact with the $\sigma$-model i.e. $\Sigma=g<\sigma>$. Clearly,
the relations in eq.(7.3) can be combined to give

$$
M_\pi^2 ={1\over{b g^2}}\Sigma^\eta (cg^2\Sigma^{2-\eta}-t)
\eqno(7.4)
$$
which, in the limit $\eta=0$ ($d=4$) reduces to the $\sigma$-model expression
(eq.(5.5)). In particular, at the critical point, $t=0$, $M_\pi^2\sim
\Sigma^2$. After recognizing that $\beta=\nu$ in this model, we recover
eq.(5.8). Similarly, in the symmetric phase, near the origin and
for fixed $t<0$, we obtain $M_\pi^2\sim |t|(\Sigma^2)^{\eta/2}$ which is
eq.(5.11). Thus,
all the general features of this plot are clear from eq.(7.4): linear
behavior in the broken phase, concavity in the symmetric
phase and the vanishing of the pion mass in both phases. The last point is
especially interesting considering the discussion of this problem in sect.5.
One can illustrate this point further by calculating the pion propagator
in the chiral limit in symmetric phase. Its form is given by  [4]
%[HKK]

$$
D_\pi^{-1}(k^2)= -t+ Ck^{2-\eta}
\eqno(7.5)
$$
which has a pole in the complex $k^2$-plane (in Minkowski space), but gives
a power law behavior of the Euclidean correlator.

As we remarked before, the general
idea behind the universal behavior of a particular
dimensionless observable is that, given its
functional form $R=G(m/t^\Delta)$, one can invert this relation to find an RG
trajectory and express another observable as $R'=R'(R)$. We
define two mass ratios: $R_\pi =M_\pi^2/M_\sigma^2$ and
$R_F=4M_F^2/M_\sigma^2$ in terms of which the above equation becomes

$$
R_\pi=1-R_F
\eqno(7.6)
$$
In the chiral limit $R_\pi=0$ and $R_F=1$ i.e. $M_\sigma=2M_F$ in the broken
phase, whereas in the symmetric phase $R_\pi\to 1$ and $R_F\to 0$ as $m\to 0$.
The curve $R_\pi$ versus $R_F$ is universal. The prediction of
the four fermi theory (eq.7.6) is a straight line with unit slope.
All the points on the curve that lie below $R_\pi=1/\delta$ belong to the
broken phase and those above to the symmetric phase.
All the physical points lie on the straight line eq.(7.6)
. Clearly, the naive heavy quark limit would require $R_\pi=R_F=1$
which is completely missing from the plot. The points below the line
$R_\pi=1/\delta$ are in the broken phase and at the critical
point $R_F=1-1/\delta$.
These two values correspond to the $m\to\infty$ limit. Explicit
calculation [13] gives for $R_\pi=G(m/t^\Delta)$

$$
{m\over{t^\Delta}}={{4b}\over{g^{2\beta}}}\Biggl(
{
{R_\pi^{1/\beta}(1-R_\pi)}\over
{(c(1-R_\pi)-4bR_\pi)^\delta}
}\Biggr)^\beta
\eqno(7.7)
$$
with $b,c$ defined as before and satisfy $4b/c=2-\epsilon=1/\beta$. Clearly,
for large $m$, the denominator on the right hand side vanishes giving

$$
R_\pi\to{c\over{c+4b}}={1\over\delta}.
\eqno(7.8)
$$

In the linear $\sigma$-model eq.(7.6) is slightly modified. The masses are
given by

$$
M_\pi^2=-t+{\lambda\over 6}v^2,\,\,\,\,\,\,
M_\sigma^2=-t+{\lambda\over 2}v^2,\,\,\,\,\,\,
M_F^2=g^2 v^2
\eqno(7.9)
$$
where $t$ is the curvature of the scalar potential ($t>0$ corresponding to
the broken phase) and $v=<\sigma>$. The equation of state is
$h=vM_\pi^2$. One can obtain the relation between the mass ratios from
eq.(7.9).
The analogue of  eq.(7.6) in this case is

$$
R_\pi=1-{\lambda\over{12g^2}}R_F
\eqno(7.10)
$$
Here, the universal curve is a straight line again, but with different
slope determined by the ratio of the coupling constants. For a fixed value of
$\lambda/g^2$ a different value of $R_F$ emerges in the chiral limit. In
that sense four-fermi is a special case of the $\sigma$ model
(as is well known) -- for $g^2=12\lambda$ they describe the same low energy
physics. Since
both mass ratios are low energy quantities, the slope should be in fact a ratio
of renormalized couplings (of course, this is not visible in the MF treatment).
In other words, the curve $R_\pi=R_\pi(R_F)$
should have no knowledge of the bare
parameters. In the four-fermi theories the renormalized couplings
are independent of the bare four-fermi coupling $G$
and the ratio is simply one.
The reason behind $M_\sigma=2M_F$ is the relative magnitude of $\lambda$
and $g$. From eq.(7.9) we see that the magnitude of $\lambda$ determines the
size of the mass scale $v$ (and meson masses),
whereas $g$ gives the magnitude of the fermion mass in units of $v$.
In four-fermi theory, unlike the $\sigma$-model, mesons are composite
rather then elementary. The compositeness condition forces the ratio of the
couplings to be such that the $\sigma$ is a true bound state i.e.
$M_\sigma\le 2M_F$. This requirement places a bound on the ratio $g^2/\lambda$
and is
intimately related to the compositeness of the $\sigma$ and $\pi$. If one
looks closely at the two couplings $\lambda$ and $g$, two things become
apparent. They appear in the lagrangian as $g\sigma\bar\psi\psi$ and
$\lambda \sigma^4$. If the $\sigma$ is composite, then {\it it renormalizes
the same way as} $\bar\psi\psi$. Therefore, compositeness requires that
$g^2$ and $\lambda$ renormalize the same way. So, if $\lambda_R=Z\lambda$,
then $g^2_R=Zg^2$ and $\lambda_R/g^2_R=\lambda/g^2$. Thus, radiative
corrections cancel and the slope of the plot
is independent of them. This, in fact, is what must
happen if the scalars are composites because the wavefunction renormalization
vanishes. If one accepts the democratic principle that, in an
interacting theory, everybody has to interact with everybody else
(unless there are selection rules that forbid it),
then in order for both couplings
to be non-zero, they have to renormalize the same way. In the $\sigma$-model
scalars are pointlike and the ratio $g^2/\lambda$ can assume any value.

In the $\sigma$-model, the
ratios are not constrained a priori
since all the degrees of freedom are pointlike. Nevertheless,
the large-$m$ behavior is consistent with that of the four-fermi model
indicating
that compositness is not the crucial ingredient here.
Rather, it is chiral symmetry and the scaling of the
mass ratios. In terms of the bare parameters, $t,h$, the mass ratio reads

$$
{h\over{t^{3/2}}}=\sqrt{
{{24}\over\lambda}
{ {R_\pi^2(1-R_\pi)}\over{(1-3R_\pi)^3}  }
}
\eqno(7.11)
$$
which, apart from the different critical exponents,
is the same as eq.(7.7) obtained for the four-fermi
theory. From
here it follows that, in the $h\to\infty$ limit, $R_\pi\to 1/3=
1/\delta$ as it should.

It has been argued many times in the literature that
either the $\sigma$ model or the
four-fermi model can be used to study the low energy regime of QCD.
This is correct provided the bare masses are sufficiently small. These models
are based on chiral symmetry and should be representative of QCD as
long as the bound states that are studied are collective. Neither model
is capable of describing atomic quarkonia like charmonium etc.
In the heavy quark limit the
predictions of these models should be qualitatively
different from QCD. Whether or not they are suitable
for the strange quark sector needs further investigation.
It is interesting to note that an upper bound on the
pseudo-Goldstone mass
follows from the scaling plot eq.(7.10). In the broken
phase, it is easy to see, that the following inequality holds

$$
M_\pi \leq {2\over{\sqrt{\delta-1}}}M_F
\eqno(7.12)
$$
In four dimension ($\delta=3$) gives $M_\pi\leq \sqrt{2} M_F$.
This bound is universal and in four dimensions holds to all orders in $1/N$
since the exponent $\delta$ does not receive any corrections.
The fact that an upper bound,
$m\to \infty$, on the $\pi-\sigma$ mass ratio is given by $1/\delta \leq 1$,
shows that an explanation of this splitting as an effect of the standard
spin-orbit interaction (\`{a} la atomic models)
is problematic simply because the splitting
survives even in the infinite mass limit.

\vskip10truemm

\noindent
{8. CONCLUSIONS}
\vskip5truemm

In this paper we have illustrated how correlation length scaling and chiral
symmetry together constrain and simplify the universal features of chiral
symmetry breaking phase transitions. Perhaps the most useful results were
1. the explicit formula for $R_\pi$ in terms of the universal function $f$,
2. the relation between the pion mass, the chiral condensate, and
the critical index $\eta$, and 3.
the simple properties of the universal function f itself. Within the context
of the analysis of simulation data, each observation 1.-3. should lead to
independent determinations of critical couplings and indices. Consistent
results from all the methods should comprise convincing evidence that one
has found universal features in the model of interest.

We have chosen to illustrate these ideas in several simple models such as
mean field theory and an ultra-violet fixed point with power-law
singularities. Other examples can also be considered and the general
approach of this paper can be applied. In ref.[22], for example, we considered
both a fixed point and a logarithmically trivial sigma model to describe the
data of four flavor lattice QED. Although both models were able to fit the
equation of state, only the fixed point model gave a consistent description
of the spectroscopy data for $R_\pi$. This exercise illustrates nicely how
the use of PCAC and scaling in unison can help lead us to the correct
theoretical scenario even when a comprehensive microscopic theory of a
critical point is missing.

We look forward to developing the ideas of this paper further and applying
them to other presently mysterious chiral phase transitions such as QCD at
nonzero chemical potential and four fermi models with continuous chiral
symmetries and few flavors in less than four dimensions.

\vskip1truecm

\noindent
{ACKNOWLEDGEMENTS}
\vskip5truemm

This work is supported in part by the National Science
Foundation under grant NSF PHY92-00148. A. K. wishes to thank
S. J. Hands for illuminating conversations.

\break
%\vskip10truemm
\noindent
{REFERENCES}
\vskip5truemm

\noindent
[1]  D.~Toussaint, Nucl.Phys. (Proc. Suppl.) 26, (1992) 3,
and references therein.

\noindent
[2]  E.~Dagotto, in: Dynamical Simmetry Breaking,
K.~Yamawaki, Editor , (World Scientific,
Singapore, 1992) p. 189, and references therein.

\noindent
[3] K.~G.~Wilson, Phys. Rev. {\bf D7} (1973) 2911; B.~Rosenstein,
B.~J.~Warr and S.~H.~Park, Phys. Rep {\bf 205} (1991) 59.

\noindent
[4]  S.~Hands, A.~Koci\'{c} and J.~B.~Kogut, {\it Four - Fermi
theories in fewer than four dimensions}, ILL-(TH)-92-\#19
S.~Hands, A.~Koci\'{c} and J.~B.~Kogut, Phys. Lett. {\bf B273}
(1991) 111;{\it The four fermi
model in three dimensions at non-zero density and temperature},
CERN-TH.6553/92 ILL-(TH)-92\#13 hep-lat/9206024.

\noindent
[5] T.~ Appelquist, {\it Dynamical EW Symmetry Breaking}, YCTP-P23-91.

\noindent
[6]  W.~A.~Bardeen, C.~T.~Hill and M.~Lindner, Phys. Rev. {\bf D41}
(1990) 1647; V.~A.~Miransky,M.~Tanabashi and K.Yamawaki, Phys.Lett.
{\bf B221} (1989) 177.

\noindent
[7] A.~M.~Horowitz, Phys.Lett. {\bf B219} (1989) 329.

\noindent
[8] C.~Itzykson and J.-M.~Drouffe, Statistical Field Theory (Cambridge
University Press, 1989).

\noindent
[9]
A.~Z.~ Patrasinskij and V.~L.~ Pokrvoskij, Zh. Eksp. Teor. Fiz.
{\bf 46} (1964) 994 (Sov. Phys. JEPT {\bf 2} (1964) 677);
L.~P.~Kadanoff,  Physics {\bf 2} (1966) 263.

%For a review of scaling and critical phenomena see, e.g.,
%A.~Z.~ Patrasinskij and V.~L.~ Pokrovskij, Fluctuation Theory of
%Phase Transitions (Pergamon Press, Oxford 1979); L.~P.~Kadanoff
%et al., Rev.~of~Mod.~Phys.~{\bf 39} (1967) 395.
%For a recent review of scaling and critical phenomena see, e.g.
%V. Privman, P. Hohenberg and A. Aharony,
%in :  Phase Transitions and Critical Phenomena, vol. 14,
%C.~Domb and J.~Lebowitz Editors, (Academic Press, London),

\noindent
[10] G.~A.~Baker,Jr, Phys. Rev. Lett. {\bf 20} (1968) 990;
 R.~Schrader, Phys. Rev. {\bf B14} (1976) 172.

\noindent
[11] J.~Glimm and A.~Jaffe, in: Recent Developments in Gauge Theories,
G.'t Hooft et al. Editors, Plenum, New York, 1980, p.201;
G.~A.~ Baker, Jr,  Phys. Rev. {\bf B15} (1977) 1553;
G.~ A.~ Baker, Jr,  in: Phase Transitions, Carg\`ese 1980,
M.~ L\'evy, J.-C.~Le~Guillou and J.~Zinn-Justin Editors, (Plenum,
New York), p. 137.

\noindent
[12] B.~D.~Josephson, Proc. Phys. Soc. {\bf 92} (1967) 269, 276.

\noindent
[13]A.~Koci\'{c}, Phys. Lett. {\bf B281} (1992) 309.

\noindent
[14] S.~J.~Hands, A.~Koci\'{c}, J.~B.~Kogut, R.~L.~Renken, D.~K.~
Sinclair and K.~C.~Wang, {\it Spectroscopy, Equation of State and
Monopole Percolation in Lattice QED with two flavors},
CERN-TH 6609/92 ILL-(TH)-92\#16 hep-lat/9208021.

\noindent
[15] G.~Boyd, J.~Fingberg, F.~Karsch,
L.~Kakkainen and B.~Petersson, Nucl. Phys. {\bf B376} (1992) 199.

\noindent
[16] R.~B.~Griffiths, Phys. Rev. {\bf 158} (1967) 176.

\noindent
[17] S.~Weinberg, Phys. Rev. {\bf 130} (1963) 776.

\noindent
[18] A.~G.~Cohen and H.~Georgi, Nucl. Phys. {\bf B314} (1989) 7.

\noindent
[19] V. Gusynin, V. Kushnir and V. Miransky,
PL {\bf B213} (1988) 177;
W.~A.~Bardeen, C.~N.~Leung and S.~T.~Love, Nucl. Phys. {\bf B323}
(1989) 493, and references therein .

\noindent
[20] K.-I.~Kondo,H.~Mino and K.~Yamawaki, Phys. Rev {\bf D39}
(1989) 2430.

\noindent
[21] K.~C.~ Bowler et al., Phys Lett {\bf 162B} (1985) 354.

\noindent
[22]  A.~Koci\'{c}, J.~B.~Kogut and K.~C.~Wang, {\it
Monopole Percolation and the Universality Class of the
Chiral Transition in Four Flavor Noncompact Lattice QED},
ILL-(TH)-92\#17

\break

\noindent
{FIGURE CAPTIONS}
\vskip5truemm

\noindent
1. $R_\pi (t,m)$ as a function of $m$ at fixed $t$ in the symmetric and
broken phase. The horizontal line corresponds to the critical point $t=0$.

\noindent
2. Typical behaviour of the universal ratio $f(x)/f(0)$  as a function of
$(x/x_0)$, $x$ being the scaled variable  $ t/<\bar\psi\psi>^{1/\beta}$.

\noindent
3. The linear dependence of $M^2_\pi$ on $<\bar\psi\psi>^2$ for a mean field
theory.

\noindent
4. Expected behaviour of $M^2_\pi$ versus $<\bar\psi\psi>^2$ in a theory
with anomalous dimension.

\end